# Mapping of functionalized regions on carbon nanotubes by scanning tunneling microscopy

P. Nemes – Incze[*,†], Z. Kónya[‡], I. Kiricsi[‡], Á. Pekker[§], Z.E. Horváth[†], K. Kamarás[§], L.P. Biró[†]

**Abstract**

Scanning tunneling microscopy (STM) gives us the opportunity to map the surface of functionalized carbon nanotubes in an energy resolved manner and with atomic precision. But this potential is largely untapped, mainly due to sample stability issues which inhibit reliable measurements. Here we present a simple and straightforward solution that makes away with this difficulty, by incorporating the functionalized multiwalled carbon nanotubes (MWCNT) into a few layer graphene – nanotube composite. This enabled us to measure energy resolved tunneling conductance maps on the nanotubes, which shed light on the level of doping, charge transfer between tube and functional groups and the dependence of defect creation or functionalization on crystallographic orientation.

In the decades following their discovery, the chemistry of carbon nanotubes (CNTs) has grown into a mature field of research[1,2,3]. Sidewall functionalization of CNTs has turned out to be a versatile tool to change their electronic properties[1,4,5,6,7] and the ways in which the tubes interact with their environment[8,9,10,11]. This broadens considerably their applicability, from sensors and medical uses[2,10,12,13] to composite materials[8]. In all such cases we need a clear picture of the changes

[*] Corresponding author email: nemes@mfa.kfki.hu
[†] Research Institute for Technical Physics and Materials Science, H-1525, PO. Box 49 Budapest, Hungary
[‡] Department of Applied and Environmental Chemistry, University of Szeged, Rerrich Béla sqr. 1, H-6720 Szeged, Hungary
[§] Research Institute for Solid State Physics and Optics, Hungarian Academy of Sciences, H-1525, PO Box 49 Budapest, Hungary

functionalization induces in the nanotube, like: surface topography, band structure, doping, etc. to be able to refine these applications and to better understand the physical and chemical processes involved in functionalization. Changes to the nanotube electronic structure are usually investigated via spectroscopic methods, which are a global probe of CNT properties[6]. Complementary to these methods are scanning probe methods, namely scanning tunneling microscopy (STM) and atomic force microscopy (AFM), which provide spatially resolved information down to the atomic scale. STM is one of the most important tools in nanoscale science and has been the source of landmark findings in the study of carbon nanotubes. Using STM, Wildöer et al. gave a first account of the relation between CNT wrapping direction and their predicted band structure[14], while Lemay et al. were able to map individual wavefunctions of CNTs[15]. These breakthroughs were possible because the tunneling current in STM can be measured simultaneously as a function of both energy and location on the sample. One way to achieve this is by using the so called Current Imaging Tunneling Spectroscopy or CITS mode, during which in each pixel of the CITS image the scanning STM tip acquires a tunneling current vs. voltage (STS) curve and stores it together with the topographic information. Using CITS image we can obtain local tunneling conductance maps of the sample by plotting the derivative of the current (dI/dV) as a function of the tip – sample bias voltage value. Among other examples, it has been applied in mapping the local photovoltaic performance of organic solar cells[16] and to map the charge inhomogeneity of $SiO_2$ supported graphene[17]. However, the full potential of STM as a tool in mapping the electronic properties of functionalized CNTs has not been achieved yet. Direct STM and CITS measurements could yield more insight into the extent and nature of sidewall functionalization as opposed to other methods of visualizing functionalized sites[18]. One key bottleneck in achieving this is sample stability.

In the case of STM, as with any high resolution microscopy tool, good mechanical stability of the sample is of paramount importance to obtain reliable measurements. Because of the presence of functional groups on the tube surface the contact area with the substrate decreases and this leads to a weakening of the overall force (ex. van der Waals) which anchors the tubes to the support. The metal tip of the STM can exert considerable forces, mainly electrostatic in nature, on the tubes[19],[20] and it can move the CNTs in the fast scan direction and can effectively "clean out" the scan area of the tubes we would want to investigate (Fig. 1). This is illustrated in Fig. 1b, where we show an AFM image of a 20x20 µm$^2$ area which was previously scanned by STM. After only 3 passes over the same area, practically all of the CNTs have been brushed to the sides. The inset in Fig. 1 shows how, during scanning of the STM tip, a nanotube can move to the side, then after a few scan lines the tube moves back to its original position. We can also find evidence of such tube movement in the literature[18,21,22,23]. While this problem seems to be less critical in the case of lightly functionalized tubes[24] and/or if the measurement is performed in UHV[25,26,27,28] it makes imaging such CNTs a challenge. One way around this problem is to immobilize the nanotubes, for example by binding the tubes to the substrate via strong chemical bonds. The work of Zhang et al. is a good example of this[22], where they have used Au – S chemistry to anchor the thiol or thiophene terminated functional groups on the nanotubes to the Au substrate, effectively immobilizing them. While fruitful, this method is not applicable in all cases because of the need to have chemical binding between the nanotubes and substrate. Furthermore, the chemical bond may perturb the tube electronic structure[22]. Here we present a method to immobilize CNTs on a graphitic support, which provides a sample that is stable enough to allow reproducible CITS measurements and is a well studied CNT – support system.

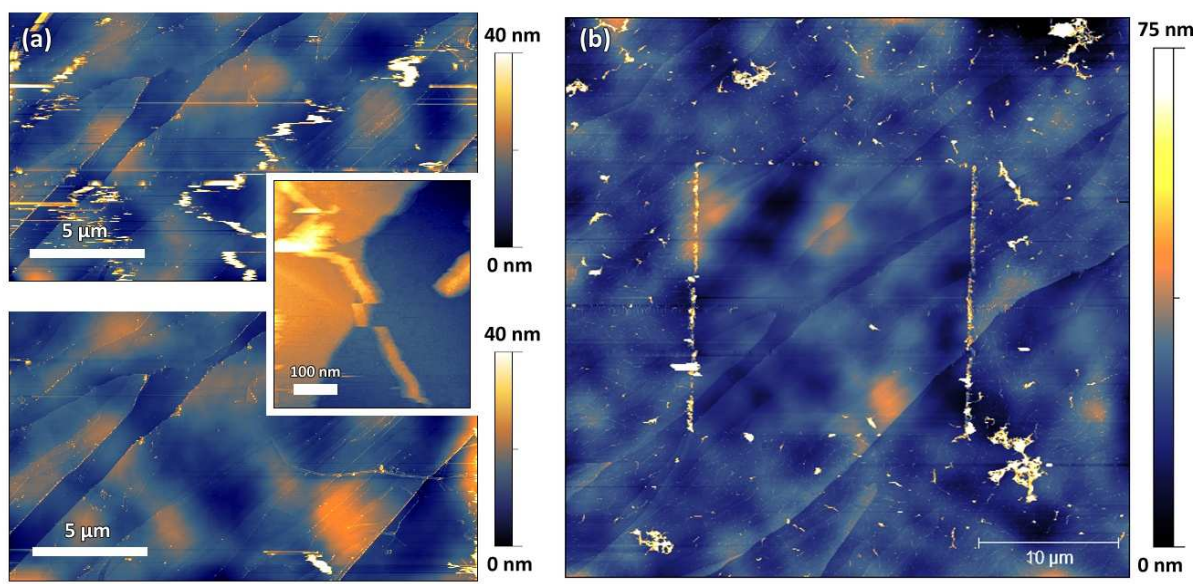

**Figure 1.** (a) Successive STM images (top and bottom) of a region on HOPG containing functionalized CNTs show that CNTs can be swept away by the STM tip. Extensive movement of the CNTs can be observed on the top image. The nanotube movement is also evident in the inset, where a single CNT shifts position during scanning and shifts back to its original position after a few scan lines. (b) Tapping mode AFM image of functionalized CNTs on a HOPG surface measured in a region where the STM tip was previously scanned (Fig. 1a). The AFM image is taken after the 3$^{rd}$ STM scan. STM scanning was performed using 300 pA current setpoint and 100 mV bias voltage. The fast scan direction is horizontal.

**Results and Discussion**

Multiwalled CNTs used in this study have been prepared by chemical vapor deposition of acetylene over alumina supported Co/Fe catalyst[29]. The functionalization of the nanotubes was done in a three step process reported elsewhere[30]. Briefly, in the first step the pristine CNTs were treated in $H_2SO_4$/$HNO_3$ (3:1) mixture for 24 h at room temperature. Such a treatment attaches –OH and –COOH groups to the CNT outer wall, which were transformed into –COCl groups in a second step involving stirring overnight in $SOCl_2$. In a final step the as obtained sample was stirred in diaminopropane for 24h. The diaminopropane can connect the functional groups on the nanotubes, which results in covalently interconnected MWCNTs as illustrated in Fig. 2a and the TEM image shown in Fig. 3c. Functionalization increases the tube surface defect concentration, as observed in the Raman spectra through an increase in the D peak intensity relative to the G peak ($I_D$/$I_G$)[31] (Fig. 2b). Attenuated Total Reflection – IR (ATR – IR) spectra of the pristine sample shows some signs of OH groups, which may be due to isopropanol residues in the CNT suspension or may be the sign of mild oxidation of the pristine sample. The spectra of the functionalized samples shows a peak corresponding to the -NH group. However, the presence of carboxyl can still be recognized, which indicates that the conversion of –COCl groups into amide was not complete. Such cross-linked carbon nanotubes are an interesting avenue in improving the mechanical and electrical transport properties of nanotube networks[32,33] and functionalization schemes similar to our own have been studied[6,34]. Here we report for the first time atomic resolution STM and CITS measurements on such CNTs.

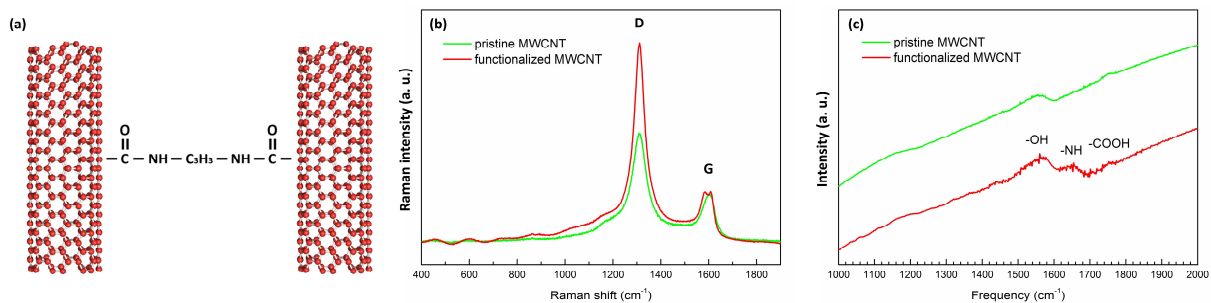

**Figure 2.** (a) MWCNTs interconnected by diaminopropane. (b) Raman spectra of the starting CNT material and of the functionalized sample. The spectra are normalized to the G peak intensity. (c) ATR-IR spectra shows peaks corresponding to –OH and –COOH groups in both the pristine and functionalized samples and a signature of –NH in the functionalized sample.

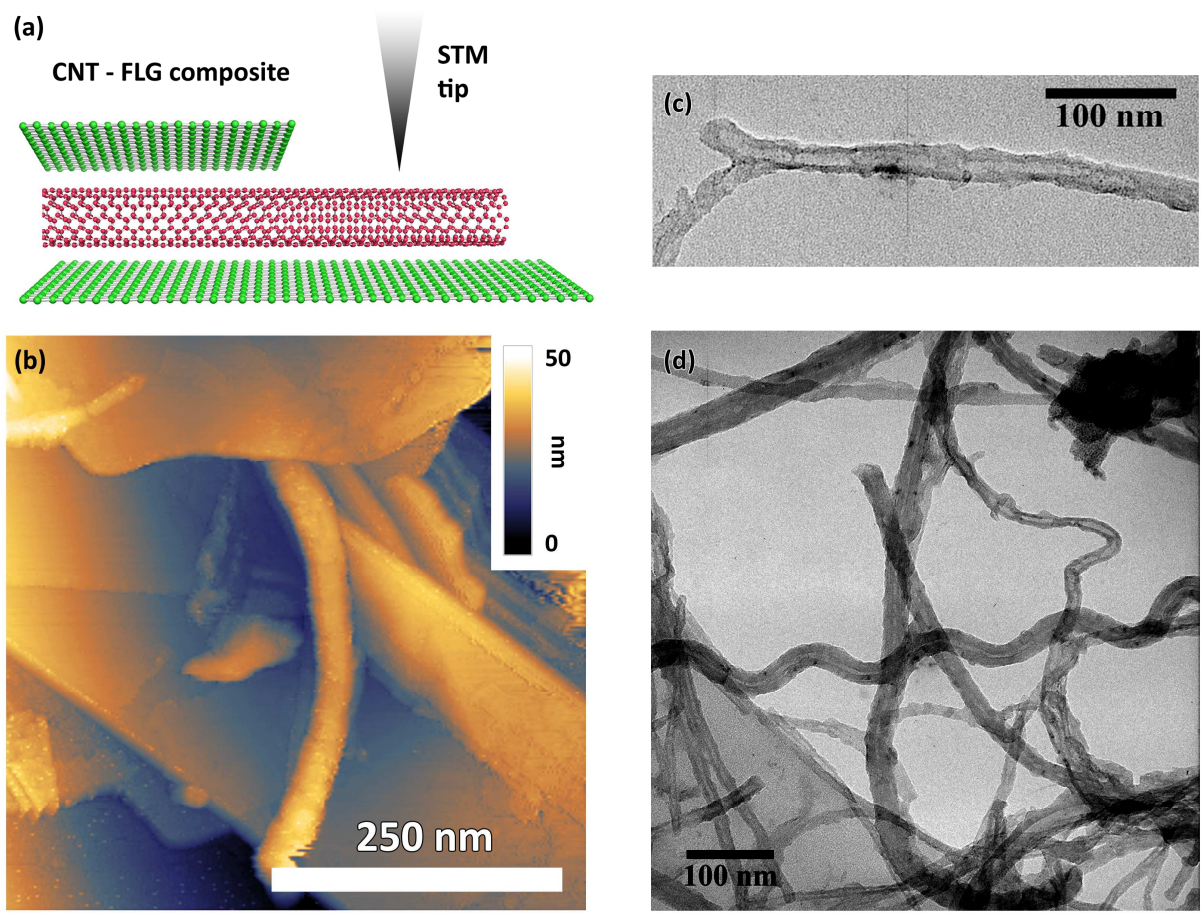

**Figure 3.** (a) Illustration of the CNT – FLG composite and the STM measurement. (b) STM image of a functionalized CNT sticking out from beneath a FLG layer, also seen in Fig. 4d. (c) and (d) TEM images of the functionalized CNTs. The surface of the tubes is irregular and some tubes are interconnected as seen in (c).

The immobilization of the functionalized CNTs is done by sandwiching them between few layer graphite (FLG) flakes in a CNT – FLG composite, i.e. a buckypaper (Fig 3a, b). The nano-composite has regions on its surface where functionalized CNTs stick out from beneath a FLG layer we show an STM image of such a region in Fig. 3b. In places like this, the movement of the nanotube is restricted by the clamping FLG layer and provides a good sample for stable and reproducible STM and CITS measurements. This particular composite is advantageous, because graphite and MWCNTs have nearly the same work function[35], thus no significant doping by the substrate is expected.

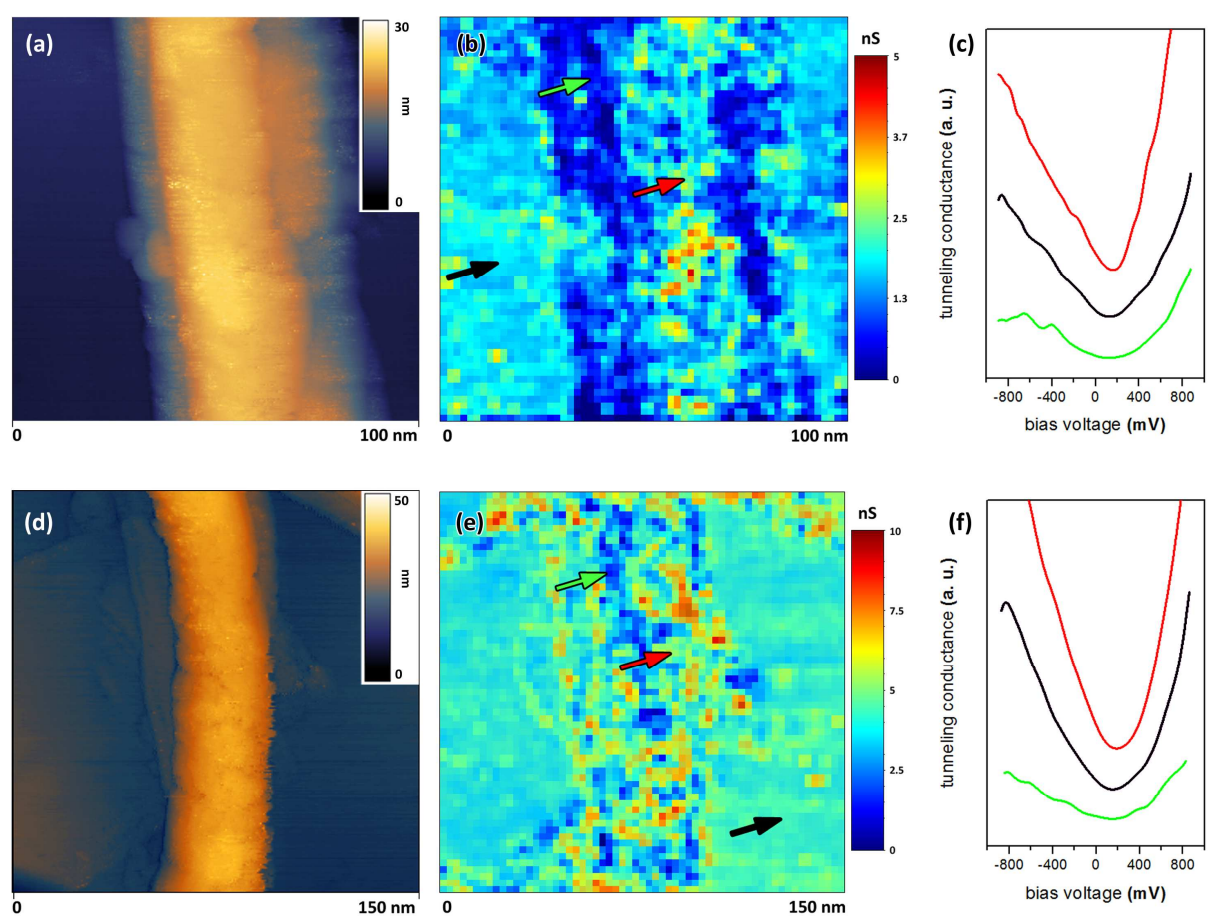

**Figure 4.** (a), (d) STM images and (b), (e) tunneling conductance maps of functionalized nanotubes extending from below a FLG clamping layer (conductance units are in nano Siemens). Figure (a) shows a bundle of 3 nanotubes, which are possibly cross linked by functional groups. The tunneling conductance maps are displayed at -160 mV bias voltage. In (c) and (f) we show dI/dV spectra from the regions marked by arrows of the same color in (b) and (e) respectively. The spectra are displaced vertically for ease of comparison.

STM images of a bundle of three functionalized nanotubes and a single tube can be seen in Fig. 4a and Fig. 4d, while in Fig. 4 b,e we present a map of the simultaneously acquired local tunneling conductance on these tubes. We obtained this map from the CITS image by taking the first derivative of the STS curves and displaying the conductance value at -160 mV bias voltage. We can observe on the same image the different conductance of the graphite support and of the nanotubes. Such dI/dV tunneling conductance maps are qualitatively proportional to the nanotube local density of states (LDOS). At a first glance the dI/dV map on the nanotube surface shows regions of high and low conductance. Comparing the dI/dV curves from such selected regions we can see some similarities and some differences in the spectra. The curves from the high conductance regions (red arrow and curve) and the substrate (black arrow and curve) are similar and have a typical graphitic shape, with monotonically increasing conductance that is symmetrical with respect to the LDOS minimum. Measuring a graphitic dI/dV signal on multiwalled CNTs is a reasonable expectation, because at the large tube diameters in this sample (~15 nm) the Van Hove singularities are very close together and get smoothed out due to the spread of the Fermi function at 300K and due to disorder effects[36]. The minimum of LDOS, which occurs because of the vanishing electron states at the K points, is shifted slightly to positive tip – sample bias voltages. This means that the chemical potential is shifted away from its equilibrium position, i.e. that the sample is p-doped. It has been shown that nitric and sulfuric acid can have a strong, p doping effect on CNTs[37,38]. In our case, acid molecules adsorbed on the nanotube surface are a result of the purification treatment. These molecules donate a positive charge to the carbon lattice, which becomes delocalized over the π electron system. Additional causes for the p doping may be the result of measuring under ambient conditions, during which oxygen or oxygen containing adsorbents can be present on the composite buckypaper (water from ambient, isopropanol from the CNT suspension)[39]. These processes lead to a "global" p doping of the nanotubes and has been measured previously by optical reflectivity and thermopower measurements [37,38]. But the true power of CITS measurements comes from the local information it provides. Thus, examining the conductance maps further we can see regions with lower conductance than the support or other regions on the tubes (green arrow and curve). dI/dV curves in these regions have a much less well defined minimum and are highly asymmetrical in places. We attributed these types of curves to defective and/or intensely functionalized regions of the CNTs, where the low conductance values arise from the destruction of the sp$^2$ π electron lattice[40]. During functionalization the sp$^2$ carbon atoms change to sp$^3$, thus reducing the number of delocalized π electrons. In the case of monovalent functional groups, such as in our case, theory predicts the existence of well defined impurity states near the Fermi level[4], but we were unable to find the signature of such states in the dI/dV curves taken above functionalized sites, contrary to other systems we have studied[41]. Again, this is a sign that in these regions the surface cannot be treated as graphitic, the functionalization does not perturb but rather completely alter the graphite lattice. This conclusion is also supported by the TEM images, where we can see that the graphitic layers of the tubes are damaged.

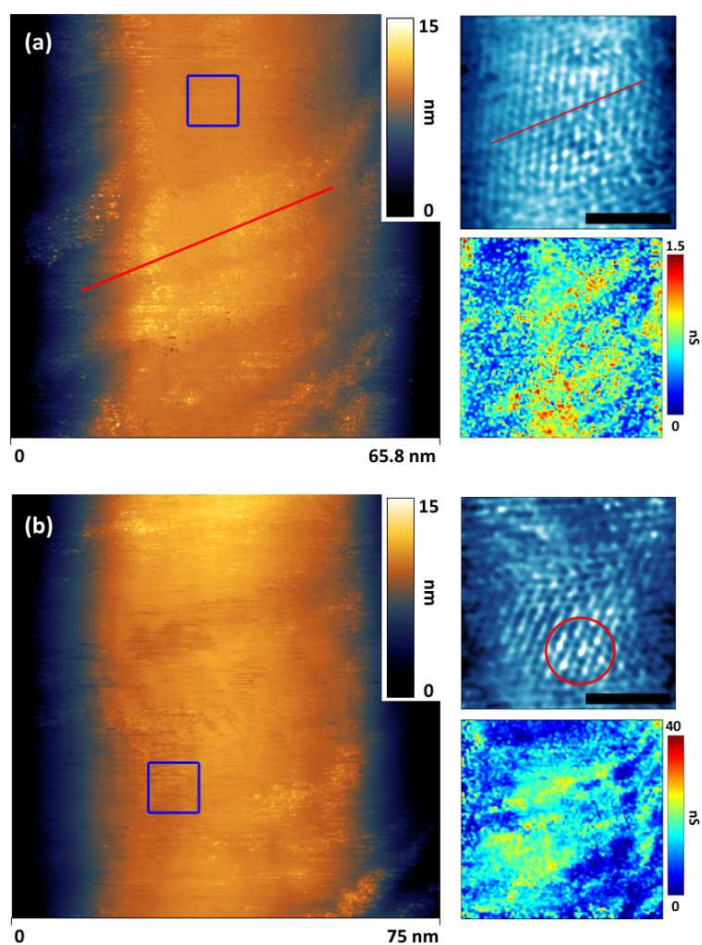

**Figure 5.** STM images of two sections of a functionalized CNT shown in Fig 4d. Zoomed in, lattice resolved images were obtained in the regions marked by blue squares. Tunneling conductance maps were obtained simultaneously and are displayed at -225 mV bias voltage. Functionalized regions show up as protrusions on the nanotube surface and as low

conductance areas in the conductance maps. Red lines mark the zigzag direction on the atomic resolution image. The functionalized regions follow the zigzag direction. The superstructure patterns on the atomic lattice are a sign of defect scattering (marked by circle).

The STM topography of these tubes also shows a rough surface, which is typical of highly defective and functionalized CNTs, where the functional groups usually appear as protrusions on the nanotube surface[22,24]. Figure 5a,b show two distinct STM topography images of a functionalized CNT. Comparing the conductance maps with the topography, we can observe that the low conductance regions show a degree of correlation with the protrusions on the surface. Especially in Fig. 5a, bands of protrusions can be observed, along with bands of low conductance in the corresponding dI/dV map. Furthermore, it was not possible to obtain atomic resolution images on the low conductance regions, only on the regions with high conductance. Alongside the graphitic dI/dV curves, this suggests that the regions with high conductance are mostly free of defects and functional groups. Here the atomic lattice shows a typical superstructure pattern, a hallmark of electron scattering between two non equivalent K points in the Brillouin zone[24,28,42]. As these superstructure patterns can extend to a few nanometers from the defect site producing them[42] it is fair to say that they are due to scattering taking place on the defects created by functionalization. If we compare the atomic resolution images with the functionalized bands, we find that these follow the zigzag direction along the nanotube circumference. To date, only a few such STM studies are available which show the correlation between crystallographic direction and functionalization, for example on fluorinated CNTs[28]. These measurements provide experimental evidence that oxidation reactions on CNTs can prefer a specific crystallographic direction, as it was presumed in recent experiments which involved the unzipping of carbon nanotubes to graphene nanoribbons[43]. Although the acid treatment in our case is much milder than in the case of CNT[43] and graphene oxidation[44], our results suggest that in the specific case discussed here, which involves nitric-sulfuric acid oxidation, this direction is the zigzag direction.

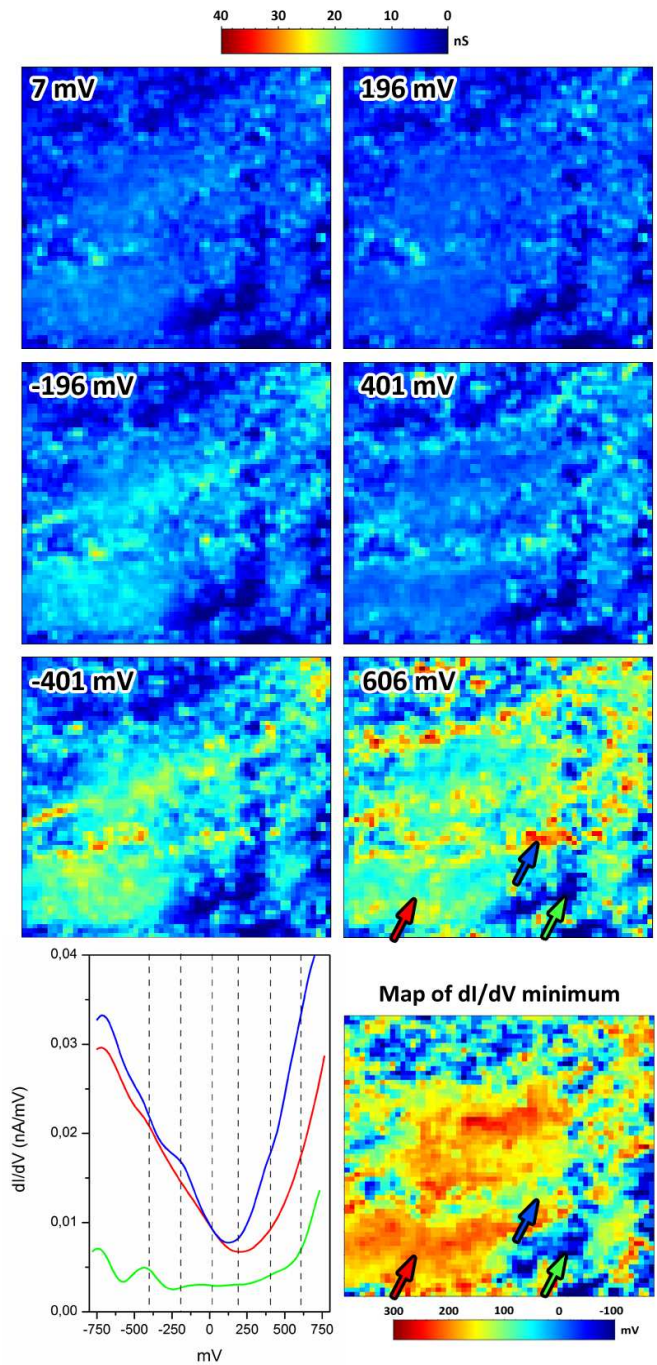

**Figure 6.** Tunneling conductance maps of the nanotube region seen in Fig. 5b, with selected dI/dV curves. The bias voltages at which the maps were plotted are marked by dashed lines. Atomic resolution was achieved in the region marked by the red arrow. The low conductance regions show similar dI/dV curves as the ones seen in Fig 4 (green). (The dI/dV maps are 75x75 nm in size). Lower right: map of the dI/dV minimum in the same region as the conductance map the bias voltage value for the dI/dV minimum is displayed on a color scale. Colored arrows show the positions where the respective dI/dV curves were taken.

Now, let us examine the dI/dV maps of one nanotube in more detail. In Fig. 6 we show six plots of the tunneling conductance map of the tube shown in Fig. 5b at different tip – sample bias voltages. These voltages have been selected in such a way, as to be symmetrical relative to the dI/dV minimum. We can distinguish three regions based on these maps. One region is where atomic resolution could be achieved (marked by a red arrow). Maps of such regions have symmetrical conductance values with respect to the minimum of dI/dV or in other words to the LDOS minimum. Such behavior is graphitic in nature, due to the symmetry in energy of the nanotube bands around the K points. As we have seen before, another region is where the overall conductance is low and no atomic resolution can be achieved. We have attributed these changes to heavily functionalized regions (marked with a green arrow). Further inspection of the maps reveals a third region, where the dI/dV maps are asymmetrical, here the tunneling conductance is higher for positive tip – sample bias voltages (blue arrow). The origin of this asymmetry is not clear, but it may be due to impurity states at functionalized or defect sites[4,41]. Another feature of this dI/dV curve is that its minimum is shifted towards a more negative sample voltage, making this region less p-doped. We can expect such behavior if there is charge transfer between the nanotube and attached molecules or radicals, which leads to a local Coulomb potential and therefore to a shift in energy of the nanotube bands. In Fig. 6 we plot a map of the LDOS minimum on the nanotube by displaying the bias voltage value for the dI/dV minimum on a color scale. From this map we can see that the mostly functional group free regions are heavily p-doped, with a Fermi level shift of around 0.25 eV. This is the global doping level of the sample, and as we discussed earlier it arises due to acid molecule adsorption and additionally it is expected if we have functionalization by electron acceptors such as carboxyl groups or other oxygen containing complexes[45]. The map of LDOS minimum correlates nicely with the conductance map showing the places of asymmetric dI/dV curves attributed to functionalized sites (blue arrows in Fig. 6). We have to mention that the interpretation of this map is not straightforward in the low conductance regions, where the $sp^2$ carbon lattice is heavily damaged and the electronic structure of the surface is unknown.

**Conclusions**

The CNT immobilization technique presented here is widely applicable to any type of functionalized nanotube and provides a well studied CNT – support system, i.e. CNT on graphite. We have shown for the first time that acid oxidation and subsequent functionalization has a preference of advancing along the zigzag crystallographic direction on large diameter multiwalled CNTs. Our measurements illustrate clearly the kind of advantage energy resolved maps can give, namely to spot sample features that are not apparent from STM topography maps and provides a local functionalization state and doping. Furthermore, this measurement technique allows certain 'in situ' studies that could not be performed otherwise, for example to examine the topography and energy resolved behavior of functionalized CNTs, while being exposed to different gas or vapor environments[10]. This may lead to a better understanding of the adsorption processes and the electronic structure variations involved in gas sensing with functionalized nanotube networks.

**Methods**

The composite presented here is a buckypaper obtained by mixing the functionalized CNTs with commercially available exfoliated graphite in isopropyl alcohol by ultrasonic stirring. The as obtained suspension was filtered through a polycarbonate membrane filter (200 nm pore size) and left to dry. Pieces from this buckypaper were cut out and placed on the STM sample holder, electrical contact to the buckypaper was made by carbon colloid paste.

A Multimode III SPM (Veeco) was operated in the CITS mode to acquire the STM images. The STS data was extracted from the raw CITS files and processed separately using Scilab (www.scilab.org) for easy manipulation and extraction of individual STS curves. All the STM images and STS curves presented here were obtained using the same Pt/Ir STM probe. Tips were mechanically cut from Pt/Ir (80:20) wire and were use for characterization only if they produced reproducible STS spectra and atomic resolution on graphite. Atomic resolution images on the nanotubes were acquired using 500 pA tunneling current and 100 mV tip – sample bias voltage. STM topography and CITS images were acquired using 300 pA tunneling current and 300 mV bias voltage.

TEM measurements were done using a Philips CM20 microscope operated at 200 keV.

**Acknowledgement**

The work was financially supported by OTKA-NKTH projects: 67793, 67851, 75813 and T049182.

One of the authors (Z.E.H.) is a grantee of a Bolyai János Scholarship.

**References**

1. Burghard, M. *Surf. Sci. Rep.* **2005**, *58*, 1-109.
2. Balasubramanian, K.; Burghard, M. *Small*. **2005**, *1*, 180-92.
3. Tasis, D.; Tagmatarchis, N.; Bianco, A.; Prato, M. *Chem. Rev.* **2006**, *106*, 1105-36.
4. Park, H.; Zhao, J.; Lu, J. P. *Nanotechnology*. **2005**, *16*, 635-638.
5. Strano, M. S.; Dyke, C. A.; Usrey, M. L.; Barone, P. W.; Allen, M. J.; Shan, H.; Kittrell, C.; Hauge, R. H.; Tour, J. M.; Smalley, R. E. *Science*. **2003**, *301*, 1519-22.
6. Hu, H.; Zhao, B.; Hamon, M. A.; Kamaras, K.; Itkis, M. E.; Haddon, R. C. *J. Am. Chem. Soc.* **2003**, *125*, 14893-900.
7. López-Bezanilla, A.; Triozon, F.; Latil, S.; Blase, X.; Roche, S. *Nano Lett.* **2009**, *9*, 940-4.
8. Sainsbury, T.; Erickson, K.; Okawa, D.; Zonte, C. S.; Fréchet, J. M.; Zettl, A. *Chem. Mater.* **2010**, *22*, 2164-2171.
9. Eitan, A.; Jiang, K.; Dukes, D.; Andrews, R.; Schadler, L. S. *Chem. Mater.* **2003**, *15*, 3198-3201.
10. Horváth, Z. E.; Koós, A. A.; Kertész, K.; Molnár, G.; Vértesy, G.; Bein, M. C.; Frigyes, T.; Mészáros, Z.; Gyulai, J.; Biró, L. P. *Appl. Phys. A*. **2008**, *93*, 495-504.
11. Klinke, C.; Hannon, J. B.; Afzali, A.; Avouris, P. *Nano Lett.* **2006**, *6*, 906-10.
12. Yun, Y.; Dong, Z.; Shanov, V.; Heineman, W.; Halsall, H.; Bhattacharya, A.; Confroti, L.; Narayan, R.; Ball, W.; Schulz, M. *Nano Today*. **2007**, *2*, 30-37.
13. Zhang, Y.; Kanungo, M.; Ho, A. J.; Freimuth, P.; van Der Lelie, D.; Chen, M.; Khamis, S. M.; Datta, S. S.; Johnson, A. T.; Misewich, J. A.; Wong, S. S. *Nano Lett.* **2007**, *7*, 3086-91.
14. Wildöer, J. W.; Venema, L. C.; Rinzler, A. G.; Smalley, R. E.; Dekker, C. *Nature*. **1998**, *391*, 59.
15. Lemay, S. G.; Janssen, J. W.; van Den Hout, M.; Mooij, M.; Bronikowski, M. J.; Willis, P. A.; Smalley, R. E.; Kouwenhoven, L. P.; Dekker, C. *Nature*. **2001**, *412*, 617-20.
16. Maturová, K.; Janssen, R. A.; Kemerink, M. *ACS nano*. **2010**, *4*, 1385-92.
17. Deshpande, A.; Bao, W.; Miao, F.; Lau, C. N.; LeRoy, B. J. *Phys. Rev. B*. **2009**, *79*, 205411.
18. Zhang, L.; Zhang, J.; Schmandt, N.; Cratty, J.; Khabashesku, V. N.; Kelly, K. F.; Barron, A. R. *Chem. Commun.* **2005**, 5429-31.
19. Curran, S.; Carroll, D. L.; Ajayan, P. M.; Redlich, P.; Roth, S.; Rühle, M.; Blau, W. *Adv. Mater.* **1998**, *10*, 311-313.
20. Biró, L. P.; Gyulai, J.; Lambin, P.; B.Nagy, J.; Lazarescu, S.; Márk, G. I.; Fonseca, A.; Surján, P. R.; Szekeres, Z.; Thiry, P. A.; Lucas, A. A. *Carbon*. **1998**, *36*, 689-696.
21. Kónya, Z.; Vesselenyi, I.; Niesz, K.; Kukovecz, A.; Demortier, A.; Fonseca, A.; Delhalle, J.; Mekhalif, Z.; B.Nagy, J.; Koós, A. A.; Osváth, Z.; Kocsonya, A.; Biró, L. P.; Kiricsi, I. *Chem. Phys. Lett.* **2002**, *360*, 429-435.
22. Zhang, J.; Zhang, L.; Khabashesku, V. N.; Barron, A. R.; Kelly, K. F. *J. Phys. Chem. C*. **2008**, *112*, 12321-12325.
23. Osváth, Z.; Koós, A. A.; Grobert, N.; Vértesy, Z.; Horváth, Z. E.; Biró, L. P. *J. Nanosci. Nanotechnol.* **2009**, *9*, 6139-6143.
24. Nemes - Incze, P.; Tapasztó, L.; Darabont, A.; Lambin, P.; Biró, L. P. *Carbon*. **2009**, *47*, 764-768.
25. Graupner, R.; Abraham, J.; Wunderlich, D.; Vencelová, A.; Lauffer, P.; Röhrl, J.; Hundhausen, M.; Ley, L.; Hirsch, A. *J. Am. Chem. Soc.* **2006**, *128*, 6683-9.

26. Bonifazi, D.; Nacci, C.; Marega, R.; Campidelli, S.; Ceballos, G.; Modesti, S.; Meneghetti, M.; Prato, M. *Nano Lett.* **2006**, *6*, 1408-14.
27. Lauffer, P.; Jung, A.; Graupner, R.; Hirsch, A.; Ley, L. *Phys. Status Solidi B*. **2006**, *243*, 3213-3216.
28. Kelly, K. F.; Chiang, I. W.; Mickelson, E. T.; Hauge, R. H.; Margrave, J. L.; Wang, X.; Scuseria, G. E.; Radloff, C.; Halas, N. J. *Chem. Phys. Lett.* **1999**, *313*, 445-450.
29. Willems, I.; Konya, Z.; Colomer, J. F.; Tendeloo, G. V.; Nagaraju, N.; Fonseca, A.; Nagy, J. B. *Chem. Phys. Lett.* **2000**, *317*, 71-76.
30. Kiricsi, I.; Konya, Z.; Niesz, K.; Koos, A. A.; Biro, L. P. In *Nanotechnology*; Vajtai, R.; Aymerich, X.; Kish, L. B.; Rubio, A. ; SPIE, 2003; Vol. 5118, pp. 280-287.
31. Rosca, I. D.; Watari, F.; Uo, M.; Akasaka, T. *Carbon*. **2005**, *43*, 3124-3131.
32. Pugno, N. M. *Mater. Today*. **2010**, *13*, 40-43.
33. Cha, S. I.; Kim, K. T.; Lee, K. H.; Mo, C. B.; Jeong, Y. J.; Hong, S. H. *Carbon*. **2008**, *46*, 482-488.
34. Chiu, P. W.; Duesberg, G. S.; Dettlaff-Weglikowska, U.; Roth, S. *Appl. Phys. Lett.* **2002**, *80*, 3811.
35. Shiraishi, M.; Ata, M. *Carbon*. **2001**, *39*, 1913-1917.
36. Hügle, S.; Egger, R. *Phys. Rev. B*. **2002**, *66*, 193311.
37. Zhou, W.; Vavro, J.; Nemes, N.; Fischer, J.; Borondics, F.; Kamarás, K.; Tanner, D. *Phys. Rev. B*. **2005**, *71*, 205423.
38. Botka, B.; Hu, H.; Niyogi, S.; Itkis, M. E.; Haddon, R. C. *Physica Status Solidi*. (in press).
39. Jhi, S. H.; Louie, S. G.; Cohen, M. *Phys. Rev. Lett.* **2000**, *85*, 1710-3.
40. Kamaras, K.; Itkis, M. E.; Hu, H.; Zhao, B.; Haddon, R. C. *Science*. **2003**, *301*, 1501.
41. Osváth, Z.; Vértesy, G.; Tapasztó, L.; Wéber, F.; Horváth, Z.; Gyulai, J.; Biró, L. P. *Phys. Rev. B*. **2005**, *72*, 045429.
42. Tapasztó, L.; Nemes-Incze, P.; Osváth, Z.; Darabont, A.; Lambin, P.; Biró, L. P. *Phys. Rev. B*. **2006**, *74*, 235422.
43. Kosynkin, D. V.; Higginbotham, A. L.; Sinitskii, A.; Lomeda, J. R.; Dimiev, A.; Price, B. K.; Tour, J. M. *Nature*. **2009**, *458*, 872-6.
44. Fujii, S.; Enoki, T. *J. Am. Chem. Soc.* **2010**, *132*, 10034-41.
45. Dukovic, G.; White, B. E.; Zhou, Z.; Wang, F.; Jockusch, S.; Steigerwald, M. L.; Heinz, T. F.; Friesner, R. A.; Turro, N. J.; Brus, L. E. *J. Am. Chem. Soc.* **2004**, *126*, 15269-76.